\documentclass[aps,prb,twocolumn,superscriptaddress]{revtex4-2}

\usepackage{amsmath,amsfonts}
\usepackage{amssymb}
\usepackage{graphicx}
\usepackage[colorlinks=true, allcolors=blue]{hyperref}
\usepackage{color}

\begin{document}

\title{Effects of the built-in electric field on free and bound excitons in a polar GaN/AlGaN/GaN based heterostructure}

\author{Loïc Méchin}\email[]{loic.mechin@uca.fr}
\affiliation{Université Clermont Auvergne, Clermont Auvergne INP, CNRS, Institut Pascal, F-63000 Clermont-Ferrand, France}
\author{François Médard}
\affiliation{Université Clermont Auvergne, Clermont Auvergne INP, CNRS, Institut Pascal, F-63000 Clermont-Ferrand, France}
\author{Joël Leymarie}
\affiliation{Université Clermont Auvergne, Clermont Auvergne INP, CNRS, Institut Pascal, F-63000 Clermont-Ferrand, France}
\author{Sophie Bouchoule}
\affiliation{Centre de Nanosciences et de Nanotechnologies, CNRS, Université Paris-Saclay, F-91120 Palaiseau, France}
\author{Blandine Alloing}
\affiliation{Université Côte d’Azur, CNRS, CRHEA, rue Bernard Gregory, Sophia Antipolis, F-06560 Valbonne, France}
\author{Jes\`us Zuñiga-Pérez}
\affiliation{Université Côte d’Azur, CNRS, CRHEA, rue Bernard Gregory, Sophia Antipolis, F-06560 Valbonne, France}
\affiliation{Majulab, International Research Laboratory IRL 3654, CNRS, Université Côte d’Azur, Sorbonne Université, National University of Singapore, Nanyang Technological University, Singapore 117543, Singapore}
\author{Pierre Disseix}
\affiliation{Université Clermont Auvergne, Clermont Auvergne INP, CNRS, Institut Pascal, F-63000 Clermont-Ferrand, France}

\date{\today}

\begin{abstract}
Low-temperature luminescence spectra reveal the presence of two independant populations of GaN excitons within a GaN/AlGaN/GaN/$\mathrm{Al_2O_3}$ heterostructure in which a thick (1.5 µm) AlGaN layer separates a thin (150 nm) top GaN layer and a thick (3.5 µm) bottom GaN layer grown on sapphire. The presence of these two spectrally-distinct families of excitons in each GaN layer of the heterostructure is demonstrated using three different experimental methods: (i) low-power µ-photoluminescence (µPL) using laser excitation sources with wavelengths above and below the AlGaN bandgap, (ii) µPL as a function of optically injected free carrier density, and (iii) quantitative numerical simulation of the µ-Reflectivity (µR). One major impact of the built-in electric field is the reduction of the excitonic lifetime in the GaN surface layer, which transitions from less than 10 ps in the presence of the built-in electric field to the bulk lifetime (90 ps) when the field is screened. This increase in the excitonic lifetime is related to the modification of the band structure in the presence of optically injected free carriers. The effect of these lifetime variations on the luminescence spectra is analyzed. Lastly, we provide an estimate of the Mott density in GaN as $n_{\mathrm{Mott}} = 4\times 10^{17}\, \mathrm{cm^{-3}}$ at 130 K, consistent with values reported in the literature and accounting for the free carrier density required to screen the electric field.
\end{abstract}


\maketitle

\section{\label{sec:level1}Introduction}

Initiated in the 1970s, the study of III-nitride materials has led to the development of LEDs \citep{Nakamura_1994,Nakamura_1995}, lasers \citep{Nakamura_1996,Nakamura_1997}, and solar-blind ultraviolet photodetectors \cite{Lim_1996}, with first LEDs operating at short wavelengths (from the near-UV to the blue) and, subsequently, extending their wavelength operation into the visible \cite{Sekiguchi_2010}. This has been possible thanks to their direct bandgap nature and their composition tunability, which enables to cover from the ultraviolet (6.12 eV for AlN \cite{Li_2003}, 3.5 eV for GaN \cite{Monemar_2002}) to the near infrared (0.65 eV for InN \cite{Wu_2002}). More recently, miniaturization of LEDs has pushed their applications beyond solid-state lighting \citep{Lee_2024,Wasisto_2019,Smith_2020,Pandey_2023}. GaN excitons exhibit a high binding energy \cite{Stepniewski_1997} and a strong oscillator strength \cite{Aoude_2008}, enabling the development of room-temperature lasers operating in the strong-coupling regime, which do not require population inversion. The strong coupling regime, where photonic and excitonic modes exchange energy coherently, was first demonstrated in vertical microcavities \citep{Weisbuch_1992,Semond_2005,Kasprzak_2006,Christmann_2008} and has been investigated in the last years within waveguide-based heterostructures \citep{Brimont_2020,Walker_2013,Walker_2015,Ciers_2017}, which simplifies the design and fabrication of the necessary heterostructures. Additionally, uncompensated dipole moments of the ionic bonds between the III and nitride elements induce a macroscopic spontaneous polarization in these materials which, combined with a piezoelectric polarization originating from strained layers, can lead to intense internal electric fields within the layers of GaN-based heterostructures. By exploiting the polarization difference between two distinct nitride materials, two-dimensional electron gas (2DEG) \citep{Ambacher_2002,Ambacher_1999} can be realized at the interface between them. Electrons confined in this gas exhibit significantly larger mobility compared to those in the bulk, and have enabled the fabrication of high electron mobility transistors (HEMTs) \citep{Khan_1993,Meneghini_2021,Shealy_2002,Sullivan_1998}, which are currently used in high-power, high-temperature electronics. While beneficial for electronic applications, these electric fields can potentially degrade the performance of exciton-based optoelectronic devices. Indeed, since excitons are composed of electrons and holes, the internal electric field in the material tends to separate these charge carriers, reducing their recombination probability and, eventually, decreasing their stability. While well-studied in quantum wells \citep{Hangleiter_1998,Leroux_1998}, the impact of this electric field on excitonic properties has been seldom explored in thicker layers. The objective of this work is to demonstrate the influence of the built-in electric field on the luminescence properties of a GaN/AlGaN/GaN heterostructure with thick layers (i.e. with no quantum confinement).

\section{\label{sec:level2}Experimental details}

The investigated sample is grown by metal-organic chemical vapor deposition (MOCVD) on a thick GaN buffer layer (3.5 µm) deposited on a $c$-plane sapphire substrate. The heterostructure of interest consists of a GaN thin layer (150 nm) on top of an AlGaN thicker layer (1.5 µm). The sample exhibits a Ga-polarity. The GaN-on-sapphire template displays a dislocation density in the order of $3\, \times 10^8 \,\mathrm{cm^{-2}}$, which is preserved by the thick AlGaN and top GaN layers. This is made possible by the use of low aluminium content AlGaN layer (8\%), enabling to sustain an almost pseudomorphic growth of the thick AlGaN, which will otherwise relax for larger thicknesses \cite{Einfeldt_2000}. Upon cooling, the difference between sapphire and nitrides coefficients of thermal expansion induce stress which is almost uniformly distributed across the structure \cite{Detchprom_1992}.\par
The difference in spontaneous and piezoelectric polarizations across the layers of the heterostructure leads to the appearance of surface charge densities at the top GaN/$\mathrm{Al_{0.08}Ga_{0.92}N}$ ($-\sigma$) and bottom $\mathrm{Al_{0.08}Ga_{0.92}N}$/GaN ($+\sigma$) interfaces. The partial compensation of these surface charges by the accumulation of free charges near the interfaces results in the formation of a two-dimensional hole gas (2DHG) at the top GaN/$\mathrm{Al_{0.08}Ga_{0.92}N}$ interface and a two-dimensional electron gas (2DEG) at the bottom $\mathrm{Al_{0.08}Ga_{0.92}N}$/GaN interface. The heterostructure is thus subjected to a significant built-in electric field within the surface GaN layer and the $\mathrm{Al_{0.08}Ga_{0.92}N}$ underlayer. The 2DHG has been observed and characterized through continuous and time-resolved spectroscopy in a previous work \cite{Mechin_2024}.\par
µPL measurements were performed using two experimental setups, each equipped with a closed-cycle helium cryostat that enable measurements at cryogenic temperatures (5.3 K). The first optical setup allows to perform time-integrated photoluminescence. Two laser sources were used: (i) a continuous wave (cw) Hübner-Photonics Cobolt 320 nm solid-state laser, and (ii) a Q-switched laser emitting at 266 nm with a pulse duration of 400 ps (21 kHz repetition rate). The high duty cycle of the Q-switched laser allows us to reach excitation densities up to $\mathrm{MW\cdot cm^{-2}}$. Thus, in combination with the cw laser, a wide range of optical intensities is covered from $10^{-3}$ to $10^6$ $\mathrm{W\cdot cm^{-2}}$. The micro-photoluminescence (µPL) signal is collected using a 100 $\times$ NUV Mitutoyo microscope objective (2 mm focal length, 0.5 numerical aperture) and a spherical lens with a focal length of 300 mm. Then, the light is dispersed by a Horiba Jobin Yvon HR1000 spectrometer (1 m focal length and 1200 gr/mm grating), and accumulated on a 1024$\times$256 pixels CCD (Charged-Coupled Device) camera. This configuration offers a spectral resolution of 0.1 nm. A Xenon lamp, emitting from 200 to 800 nm under normal incidence, was used to perform micro-reflectivity (µR) measurements.\par
The second experimental setup is used for time-resolved photoluminescence (TRPL). A Ti:sapphire laser with a pulse duration of 150 fs and a repetition rate of 76 MHz is employed to obtain second and third harmonics, at 349 nm and 266 nm respectively, of its output wavelength. Please note that 349 nm lies in between the $\mathrm{Al_{0.08}Ga_{0.92}N}$ and the GaN absorption edges, enabling us to pump the top and bottom GaN, while all the other pump sources will be completely absorbed by the top GaN and $\mathrm{Al_{0.08}Ga_{0.92}N}$ layers. Detection and analysis of the luminescence signal are performed in a similar way to the first setup. The spectrometer has a lesser spectral resolution as it is equipped with a 600 gr/mm grating and has a shorter focal length. A Hamamatsu streak camera is placed on the output plane of the spectrometer to analyse temporally the photoluminescence signal with a temporal resolution of about 10 ps in the setup configuration used in this study. The transmission of the microscope objective used for both setups requires to excite the sample from the side when using the Q-switched and Ti:sapphire lasers. A confocal excitation geometry is used with the cw laser.

\section{\label{sec:level3}Identification of excitonic transitions}
The GaN top layer of the GaN/$\mathrm{Al_{0.08}Ga_{0.92}N}$/GaN heterostructure is subject to a strong built-in electric field as discussed in our previous study \cite{Mechin_2024}. The ionization field $\mathrm{F_{ion}}$ of about 100 $\mathrm{kV\cdot cm^{-1}}$, as determined with an hydrogenoïd model \cite{Yamabe_1977}, is probably exceeded in part of the layer. Thus the presence of excitons has to be carefully demonstrated. Moreover, even for lower electric fields, effects such as broadening or shift may alter the excitonic transitions as discussed by Winzer and coworkers \cite{Winzer_2006}.\par
\begin{figure}
	\includegraphics[width=\columnwidth]{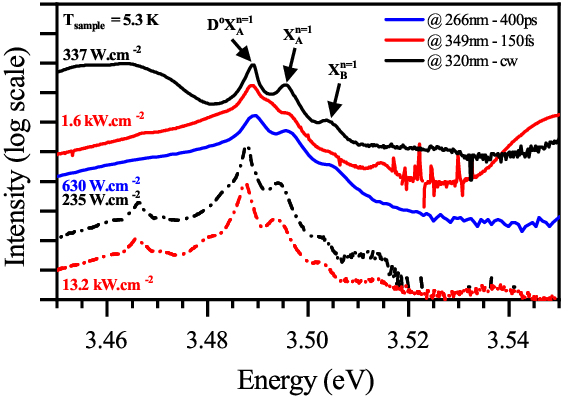}
	\caption{\label{Fig1} Low-excitation µPL spectra obtained for different laser sources: Q-switched laser at 266 nm (blue line), femtosecond Ti:sapphire laser at 349 nm (red lines), cw laser at 320 nm (black lines). The solid lines correspond to the experimental measurements performed on the GaN/$\mathrm{Al_{0.08}Ga_{0.92}N}$/GaN heterostructure. The dashed lines correspond to the experimental measurements on the thick GaN buffer layer on sapphire. The baseline of each spectrum is shifted vertically for clarity.}
\end{figure}
Figure \ref{Fig1} shows low-excitation µPL spectra, obtained using different laser pump sources, of the GaN/$\mathrm{Al_{0.08}Ga_{0.92}N}$/GaN heterostructure (solid lines). The spectra obtained with 320 nm or 266 nm, which are completely absorbed by the top GaN and by the top tens of nanometers of the $\mathrm{Al_{0.08}Ga_{0.92}N}$ layers, show excitonic transition energies shifted at energies higher than that of excitons in fully-relaxed GaN: $\mathrm{D^\circ X_A^{n=1}}=3.4890$ eV, $\mathrm{X_A^{n=1}}=3.4955$ eV, $\mathrm{X_B^{n=1}}=3.5037$ eV. This shift at high energy confirms that the top GaN layer is under a compressive strain of about -10 kbar \cite{Gil_1997}. To confirm these results, we have performed µPL on a thick GaN buffer layer (6.5 µm) grown on sapphire with the same growth conditions. The spectra obtained from the latter (dashed lines) is dominated also by excitonic transitions: free $A$ ($\mathrm{X_A^{n=1}}=3.4940$ eV) and $B$ ($\mathrm{X_B^{n=1}}=3.5025$ eV) excitons, and the $A$ exciton bound to a neutral donor ($\mathrm{D^\circ X_A^{n=1}}=3.4879$ eV). The localization energy of 6.1 meV suggests that the neutral donor is a silicon atom \cite{Monemar_2008}. At lower energy (3.4665 eV), the two-electron satellite transitions (TES) confirms this interpretation \cite{Freitas_2002}. When excitation is set at 349 nm, the main excitonic transition energies of the heterostructure remain unchanged, but a new luminescence peak appears at approximately 3.4930 eV, between the free $A$ exciton and the neutral donor-bound exciton.\par
In order to determine the layer at the origin of each of these excitonic emissions, let us note that the 150 nm-thick GaN layer of the GaN/$\mathrm{Al_{0.08}Ga_{0.92}N}$/GaN heterostructure is not sufficient to absorb all the incident light. By considering the absorption coefficient at the laser wavelength ($\mathrm{1.2\times 10^5 \, cm^{-1}}$ \cite{Muth_1997}), 16.5\% of the incident photons can cross this layer and create charge carriers (i) in the $\mathrm{Al_{0.08}Ga_{0.92}N}$ layer when excitation is set at 320 nm or 266 nm, or (ii) directly in the GaN buffer when excitation is set at 349 nm ($\mathrm{\lambda_G^{AlGaN}}=$ 339.6 nm). Since a small portion of the incident photons reaches the buffer GaN layer, this luminescence peak can be attributed to the free $A$ exciton from this layer. The small energy differences between the excitonic transitions observed here can be explained by slight variations in the strain state imposed by the $\mathrm{Al_{0.08}Ga_{0.92}N}$ on the GaN layers during the cooling of the structure.\par
\begin{figure}
	\includegraphics[width=\columnwidth]{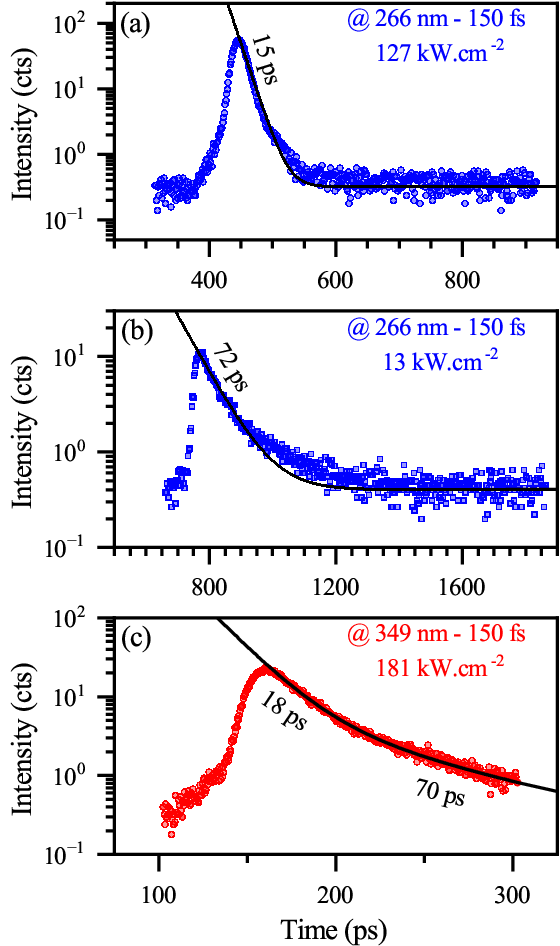}
	\caption{\label{Fig2} Time-Resolved PhotoLuminescence (TRPL) results obtained on the complete heterostructure ((a) and (c)) and on the thick GaN layer ((b)). The laser output wavelength is modulated to probe only the signal originated from GaN layers. The black lines correspond to monoexponential ((a) and (b)) and biexponential fits ((c)) of the experimental luminescence decays. The spectral range of integration was chosen to limits the contribution of bound exciton luminescence.}
\end{figure}
To confirm that both the top and the buffer GaN layers contribute to the photoluminescence, time-resolved photoluminescence (TRPL) measurements were performed. Figure \ref{Fig2}\textcolor{blue}{(a)} shows the luminescence decay of the $A$ exciton under excitation at 266 nm with a power density of 127 $\mathrm{kW\cdot cm^{-2}}$ on GaN/$\mathrm{Al_{0.08}Ga_{0.92}N}$ heterostructure. At 266 nm, the incident photons are fully absorbed within the surface GaN layer and the upper region of the $\mathrm{Al_{0.08}Ga_{0.92}N}$ layer. Thus, the signal observed  should correspond essentialy to the recombination of excitons in the 150 nm surface GaN layer. The decay was fitted using a mono-exponential model, yielding an exciton lifetime of 15 ps. Figure \ref{Fig2}\textcolor{blue}{(b)} shows the results obtained for the thick GaN buffer layer under the same conditions as before but with the excitation density reduced by a factor of 10 (13 $\mathrm{kW\cdot cm^{-2}}$) to account for the excitation conditions of the GaN buffer layer with the 349 nm excitation of the GaN/$\mathrm{Al_{0.08}Ga_{0.92}N}$/GaN heterostructure. A mono-exponential decay of $A$ exciton is again observed, and the fit yields a longer exciton lifetime of 72 ps. Finally, Figure \ref{Fig2}\textcolor{blue}{(c)} shows the time-resolved photoluminescence signal obtained for the GaN/$\mathrm{Al_{0.08}Ga_{0.92}N}$/GaN heterostructure under 349 nm excitation. In this case, a portion of the incident photons is not absorbed by the surface GaN layer and passes through the transparent $\mathrm{Al_{0.08}Ga_{0.92}N}$ layer before being absorbed by the GaN buffer layer. The experimental results show a biexponential decay of the $A$ exciton luminescence, indicating the presence of two independent excitonic populations. Fitting the experimental decay with a biexponential function yields to decay times for the two populations: 18 ps and 70 ps. The comparison between these different decays evidence that the observed time-resolved signals originate, partly, from the thin top GaN, clearly indicating the formation of excitons despite the built-in electric field. Moreover, as seen in Figure \ref{Fig1}, their signal dominates the µPL spectra.

\section{\label{sec:level4}Photoluminescence as a function of excitation optical intensity}

\begin{figure}
	\includegraphics[width=\columnwidth]{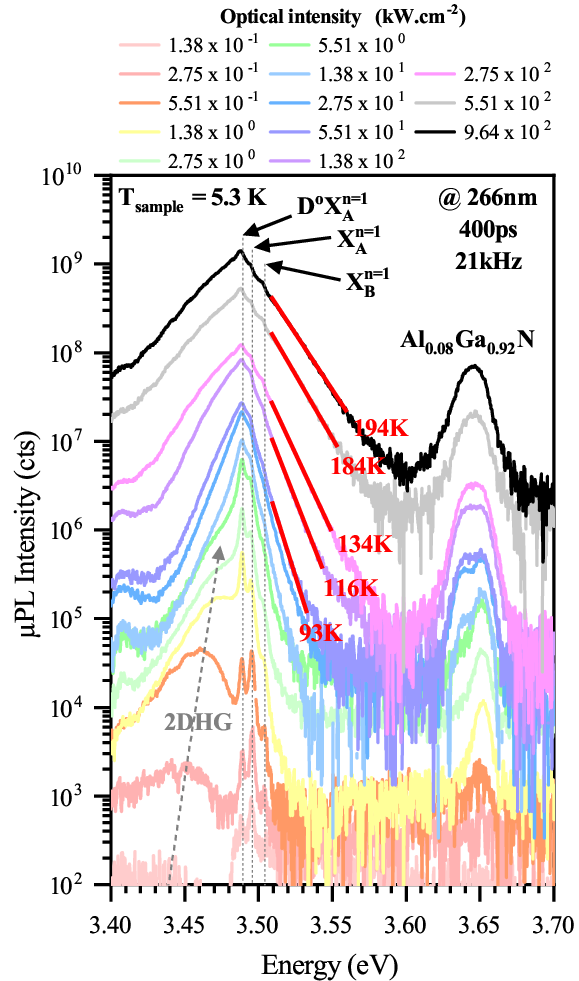}
	\caption{\label{Fig3} µPL spectra as a function of optical intensity obtained using the Q-switched laser at 266 nm. The main excitonic transitions, as well as the luminescence peaks associated with the 2DHG and $\mathrm{Al_{0.08}Ga_{0.92}N}$, are reported. The electronic temperature in the surface GaN layer is determined by analyzing the exponential decay of the high-energy side of the GaN photoluminescence intensity (red lines).}
\end{figure}
As the excitation optical intensity is increased, more free carriers are created in the GaN layer. The injected free carriers modify the electronic potential etablished across the GaN/$\mathrm{Al_{0.08}Ga_{0.92}N}$/GaN heterostructure by screening the built-in electric field, thereby modifying their own recombination processes. These changes should be observed in the luminescence spectra. In this section, we measure the luminescence spectra as function of the excitation optical intensity with an excitation wavelength of 266 nm. The figure \ref{Fig3} shows the µPL spectra obtained using the Q-switched laser at 266 nm as a function of the optical intensity. For low intensities (from 0.138 $\mathrm{kW\cdot cm^{-2}}$ to 0.551 $\mathrm{kW\cdot cm^{-2}}$), the transition associated to the 2DHG is evident, indicating the presence of a strong internal electric field (which results in the band configuration responsible for the 2DHG formation). The intensity of this electric field also affects the excitonic transitions in the top GaN layer: the luminescence peak of $\mathrm{D^\circ X_A^{n=1}}$ is weaker than that of $\mathrm{X_A^{n=1}}$. From 0.138 $\mathrm{kW\cdot cm^{-2}}$ to 13.8 $\mathrm{kW\cdot cm^{-2}}$, the 2DHG transition blueshifts and the luminescence intensity ratio between $\mathrm{X_A^{n=1}}$ and $\mathrm{D^\circ X_A^{n=1}}$ reverses, indicating a progressive screening of the electric field. At 13.8 $\mathrm{kW\cdot cm^{-2}}$, the disappearance of the 2DHG transition indicates the complete screening of the electric field by the free carriers. The appearance of a luminescence peak at 3.640 eV, close to the initial luminescence peak of $\mathrm{Al_{0.08}Ga_{0.92}N}$ (3.650 eV), confirms the screening of the electric field. The charge carriers generated at the surface of the $\mathrm{Al_{0.08}Ga_{0.92}N}$ layer in the presence of the electric field were driven (i) into the interior of the layer for electrons, and (ii) into the potential well of the 2DHG for holes. The luminescence from $\mathrm{Al_{0.08}Ga_{0.92}N}$ observed at low excitation densities corresponds to the recombination of electrons localised deep in the $\mathrm{Al_{0.08}Ga_{0.92}N}$ layer. In the absence of an electric field, electron-hole recombination more likely occurs at the surface of the $\mathrm{Al_{0.08}Ga_{0.92}N}$ layer, where the aluminum fraction may be locally lower due to interdiffusion between the $\mathrm{Al_{0.08}Ga_{0.92}N}$ and GaN layers, as observed previously \cite{Nemoz_2021}. Another possibility is the recombination of free excitons localized on neutral donors, unobservable at low excitation intensities due to the ionization of donors by the built-in electric field in the top of $\mathrm{Al_{0.08}Ga_{0.92}N}$ layer, similarly to excitonic transitions in the top GaN layer. When more free carriers are injected, the electronic temperature of GaN carriers increases (from 93 K for 55.1 $\mathrm{kW\cdot cm^{-2}}$ to 194 K for 964 $\mathrm{kW\cdot cm^{-2}}$). Despite this increase in electronic temperature and the high density of free carriers, excitonic transition peaks of GaN remain observable.

\section{\label{sec:level5}Built-in electric field effects on excitonic properties}

The experimental results indicate that the photoluminescence spectra strongly depend on the built-in electric field. To understand how the electric field influences the transitions within the heterostructure, the band diagram was calculated using \textit{nextnano} software, which self-consistently solves the one-dimensional Schrödinger and Poisson equations. For the simulations, the temperature was set to 10 K (instead of 5 K) to approximate the experimental conditions while avoiding numerical issues associated with the strong variations in Fermi-Dirac statistics near 0 K. The Fermi level pinning at the surface was fixed at 1 eV to match the Schottky barrier height measured in GaN Schottky diodes \citep{Gladysiewicz_2013, Janicki_2021}. The donor density was set to $\mathrm{N_D = 2.5 \times 10^{17}\, cm^{-3}}$, consistent with values reported in the literature for GaN layers on sapphire grown by MOVPE \cite{Koleske_2002}. A donor ionization energy of 30 meV was used, in agreement with literature values for silicon \cite{Wysmolek_2002}.\par
\begin{figure}
	\includegraphics[width=\columnwidth]{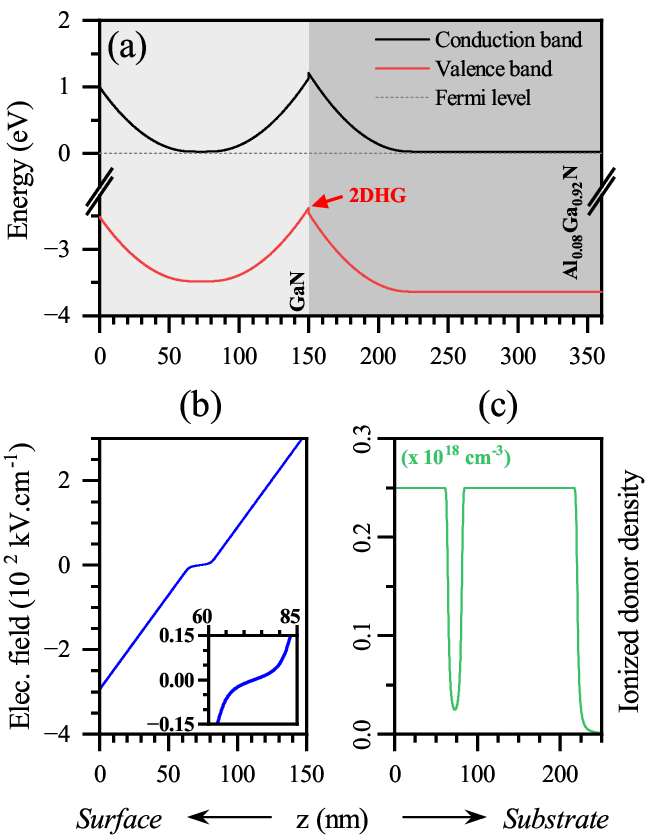}
	\caption{\label{Fig4} (a) Numerical simulation of band alignment in the structure using a self-consistent
one-dimensional resolution of the Schrödinger and Poisson equations. The simulation highlights the presence of a 2DHG at the GaN/$\mathrm{Al_{0.08}Ga_{0.92}N}$ interface. The strong built-in electric fields across the structure tend to accumulate free electrons in the middle of the top GaN layer and push the holes toward the surfaces. (b) Electric field amplitude across the top GaN layer: the electric field is extremely strong near the edges of the layer, but the inversion of the field direction defines a 15 nm thick region where the electric field is weaker. (c) Ionized donor density as a function of position in the heterostructure: the electric field is strong enough to ionize donors over a significant portion of the top GaN layer.}
\end{figure}

Figure \ref{Fig4}\textcolor{blue}{(a)} displays the calculated band diagram for the heterostructure. The simulation reveals that the bands are highly bent near the surface and at the GaN/$\mathrm{Al_{0.08}Ga_{0.92}N}$ interface. Near the sample surface, this bending is due to the pinning of the Fermi level, which depends on the electronic states available at the surface and cannot be modified experimentally in a non-destructive manner. At the GaN/$\mathrm{Al_{0.08}Ga_{0.92}N}$ interface, the band bending is caused by the polarization discontinuity across the heterostructure and is partially compensated by the accumulation of free holes in the 2DHG.\par
In the simulations presented in Figure \ref{Fig4}, the system is considered at the electrostatic equilibrium, without free charge carriers injected. Under optical injection, charge carriers are injected and modify the band structure as the electric field is partly screened. However, for a weak intensity, the carriers can be assumed to experience the calculated band structure. For weak excitation using the Q-switched laser at 266 nm, carriers are injected homogeneously into the GaN layer, with a small fraction generated in the upper part of the $\mathrm{Al_{0.08}Ga_{0.92}N}$ layer by photons not absorbed in the top GaN layer. The photogenerated carriers drift in the heterostructure depends on their charge: (i) electrons drift to the center of the layers, while (ii) holes accumulate at the interfaces (sample surface and 2DHG). Experimental results reveal that the photoluminescence signal from the 2DHG is very weak for an optical intensity of 138 $\mathrm{W\cdot cm^{-2}}$ and increases rapidly to dominate the spectrum at 551 $\mathrm{W\cdot cm^{-2}}$. Simulation results with the choosen donor concentration indicate that the potential well formed by the 2DHG lies below the Fermi level at equilibrium. Therefore, a sufficient number of holes must be injected into the structure to lower the quasi-Fermi level for holes into the potential well associated with the 2DHG, enabling optical recombination involving these confined levels. This explains why a weak or no signal from the 2DHG is observed at lowest excitation densities. In the same excitation density range, the neutral donor-bound exciton $\mathrm{D^\circ X_A^{n=1}}$ emission is less intense than that corresponding to the free exciton $\mathrm{X_A^{n=1}}$. Figures \ref{Fig4}\textcolor{blue}{(b)} and \ref{Fig4}\textcolor{blue}{(c)} respectively detail the electric field amplitude together with the ionized donor density in the GaN surface layer. The reversal of the electric field sign in the surface layer defines a region where the field is sufficiently weak to partially ionize donors but maintain free excitons. Since the $\mathrm{D^+X}$ complex is unstable in GaN \cite{Corfdir_2014}, this weak electric field explains the inversion of the luminescence intensity ratio between $\mathrm{D^\circ X_A^{n=1}}$ and $\mathrm{X_A^{n=1}}$ compared to classical spectra obtained on a GaN reference sample.\par
\begin{figure}
	\includegraphics[width=\columnwidth]{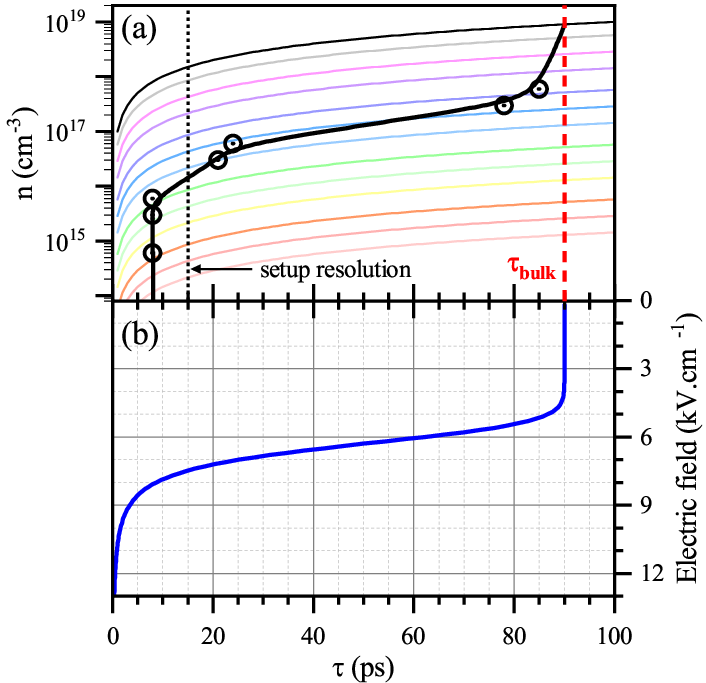}
	\caption{\label{Fig5} (a) Graphical determination of the optically injected carrier density. The colored lines represent the optically injected carrier density using the Q-switched laser at 266 nm, as a function of the carrier lifetime in the GaN layer $\tau$. Each line corresponds to a specific intensity (ranging from 138 $\mathrm{W\cdot cm^{-2}}$ (beige line) to 964 $\mathrm{kW\cdot cm^{-2}}$ (thin black line)). The black circles represent the experimental determination of the excitonic lifetime from time-resolved photoluminescence measurements, determined in reference \cite{Mechin_2024}. The thick black line is an interpolation of the experimentally measured optically injected carrier density using the femtosecond laser at 349 nm. (b) Lifetime $\tau$ of excitons as function of built-in electric field $F$ (\textcolor{blue}{eq.}\ref{eq1}). We see that a small electric field is sufficient to decrease sharply the excitonic lifetime.}
\end{figure}
When the optical intensity increases, the number of free carriers injected into the structure rises, leading to a modification of the band structure. Dynamically simulating the band edge to model the effect of optically injected charge carriers is challenging and has not been realized in this work. To better understand how the built-in electric field evolves with increasing carrier injection density, another approach is to analyze the evolution of the exciton lifetime in the surface GaN layer. The dissociation rate of a free exciton in an electric field $F$ can be described as the ionization probability of a hydrogen atom \cite{Yamabe_1977}:
\begin{equation}\label{eq1}
\frac{1}{\tau_d}=\omega \frac{4F_{ion}}{F}\exp\left\lbrace -\frac{2F_{ion}}{3F}\right\rbrace ,
\end{equation}  
where $\omega=R_y /\hbar$ is the frequency corresponding to the exciton binding energy $R_y =$ 25 meV in GaN, and $F_{ion} = R_y/e_0a_B$ is the field capable of a voltage drop of one Rydberg \cite{Blossey_1970}, expressed as a function of the Bohr radius $a_B = \hbar /\sqrt{2\mu R_y}\sim$ 2.8 nm. Here, $\mu = \mathrm{0.197}m_0$ is the reduced mass of the exciton \cite{Vurgaftman_2003}. Figure \ref{Fig5}\textcolor{blue}{(b)} shows the calculated exciton lifetime $\tau$ as a function of the built-in electric field $F$. The results indicate that an electric field of approximately 12 $\mathrm{kV\cdot cm^{-1}}$ is sufficient to separate the electron-hole pairs almost instantaneously. Between 9 and 4 $\mathrm{kV\cdot cm^{-1}}$, the dissociation time varies rapidly, approaching the non-radiative lifetime of the free exciton as measured by TRPL on a GaN buffer layer. To compare this simplified picture with experimental results, TRPL measurements were performed on the heterostructure to determine the evolution of the free exciton lifetime as a function of the density of optically injected carriers. We observe a rapidly increasing lifetime from the setup resolution (approximately 10 ps) up to a value close to the one measured on the buffer layer (85 ps) by just changing the excitation density by a factor 100. To know the density of injected carriers necessary to screen the built-in electric field, we have determined the number of optically injected charge carriers $n$ as a function of the pulse light intensity. We used an excitation wavelength of 266nm to avoid observing the luminescence signal of the excitonic transitions of the GaN buffer layer. The pulse duration is much shorter than the characteristic recombination time in GaN ($\tau > t_p=150$ fs). Under these pumping conditions one can considered a pulsed regime, for which the injected carrier density can be calculated as:
\begin{equation}
n=Gt_p=\frac{(1-R)It_p}{h\nu d},
\end{equation}
where $R$ is the reflectance at the vacuum/GaN interface ($\sim$ 0.03 for excitation at an incidence angle of 64° by taking into account the TM laser polarization), $I$ is the experimentally measured optical intensity in $\mathrm{W\cdot cm^{-2}}$, $h\nu$ is the energy of incident photons, and $d$ is the thickness of the material over which the light pulse is absorbed. The diffusion length of carriers in GaN is much greater \cite{Duboz_1997} than the thickness of the GaN layer considered (150 nm). Thus, a uniform injection across the layer's thickness ($d=150$ nm) is assumed. The experimental determination of $\tau$ using time-resolved photoluminescence \cite{Mechin_2024} allows us to plot the experimental points on graph \ref{Fig5}\textcolor{blue}{(a)} (black circles). It is observed that for low charge carrier densities (from $\mathrm{10^{14}\, cm^{-3}}$ to $6\times\mathrm{10^{15}\, cm^{-3}}$), the measured excitonic lifetime is shorter than the resolution of the experimental setup (approximately 10 ps). As the injected carrier density increases (from $\mathrm{10^{16}\, cm^{-3}}$ to $\mathrm{10^{18}\, cm^{-3}}$), the lifetime rapidly increases and approaches the measured bulk GaN lifetime (90 ps). These experimental observations are consistent with the theoretical model presented earlier. In terms of band structure and built-in electric field, the injection of free carriers screens the negative surface charge density at the GaN/$\mathrm{Al_{0.08}Ga_{0.92}N}$ interface through the accumulation of free holes in the 2DHG. At the surface of the sample, free carriers fill deep levels, decreasing the Fermi level pinning. These screenings are accompanied by a widening of the region within the GaN layer where the electric field is weak, as reflected by the increase in the excitonic lifetime measured by TRPL. When the injected carrier density becomes sufficiently high to completely screen the polarization-induced surface charge density, the electric field is canceled from the middle of the GaN layer to the interfaces, resulting in a flattening of the valence and conduction bands.\par
\begin{figure}
	\includegraphics[width=\columnwidth]{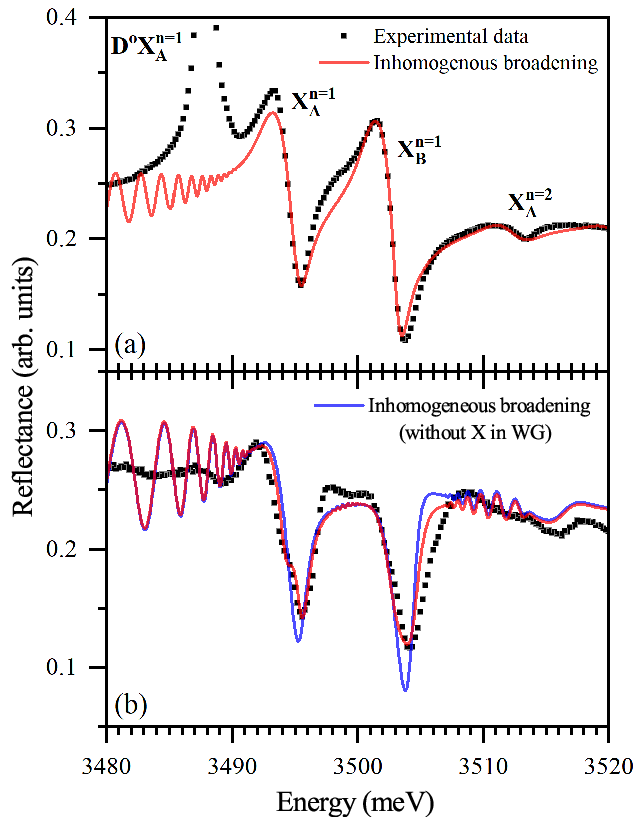}
	\caption{\label{Fig6} Experimental results of the micro-reflectivity measurements at low temperature (5.3 K) are represented by the black squares for (a) the GaN buffer on sapphire and (b) the GaN/$\mathrm{Al_{0.08}Ga_{0.92}N}$/GaN heterostructure. The observed peak at 3487.8 meV on graph (a) is not a reflectivity feature but is the luminescence peak of the exciton bound to a neutral donor. Solid lines correspond to the numerical simulation of the reflectivity using the transfer matrix formalism for a multilayer structure. The excitonic resonances are modeled by using an inhomogeneous broadening model.}
\end{figure}
Under low carrier injection conditions, the region where the electric field is sufficiently weak to maintain free excitons is probably small (approximately 15 nm for $\mathrm{N_D=2.5\times 10^{17} \, cm^{-3}}$, as calculated by the simulations). One way to validate these results is to perform µR measurements to measure the actual excitonic properties (i.e. resonance energy, oscillator strength, homogenous and inhomogenous broadenings). Figure \ref{Fig5}\textcolor{blue}{(a)} shows a µR spectrum of the thick GaN buffer layer grown on sapphire which clearly displays the two main excitonic transitions, $\mathrm{X_A^{n=1}}$ and $\mathrm{X_B^{n=1}}$ , as well as the first excited states: $\mathrm{X_A^{n=2}}$ and $\mathrm{X_B^{n=2}}$. Due to the good quality of the elaborated GaN, the photoluminescence of $\mathrm{D^\circ X_A^{n=1}}$  is super imposed in the reflectivity spectrum. The energies of the excitonic transitions are determined by simulating the reflectivity spectrum numerically. For this, we applied the transfer matrix formalism for a multilayer system, treating the first (air) and last ($\mathrm{Al_2O_3}$) layers as semi-infinite. The simulation accounts for the dispersive indices of each material, for which the contribution of excitonic transitions has been removed. This contribution of the latter is then added to the GaN dielectric function within an inhomogeneous model:
\begin{align}
\varepsilon (E)= &\varepsilon_b + \sum\limits_j \int\limits_0^{+\infty} \frac{1}{\sqrt{2\pi}\sigma_j} \frac{f_j}{x^2 - E^2 + i\gamma_j E}\\
&\times \exp\left\lbrace -\frac{(x-E_j)^2}{2\sigma_j^2}\right\rbrace dx;\nonumber
\end{align}
where $\varepsilon_b$ represents the background dielectric function. Each excitonic resonance $j$ is associated with the following physical properties: $f_j$ is the oscillator strength, $E_j$ is its energy, $\gamma_j$ and $\sigma_j$ are respectively the homogeneous and inhomogeneous broadenings of the transition. Adjustment of simulations to the experimental results permitted us to obtain the physical parameters of the excitons. The energies of the excitonic transitions determined through this model are as follows: $\mathrm{E(X_A^{n=1})}$=3494.0 meV and $\mathrm{E(X_B^{n=1})}$=3502.0 meV. The main excitonic transitions exhibit a homogeneous broadening of $\gamma_A = 0.45$ meV and $\gamma_B = 0.5$ meV, along with inhomogeneous broadenings of $\sigma_A = 0.8$ meV and $\sigma_B = 0.73$ meV. The determined oscillator strengths are $f_A = 35\,000\,\mathrm{meV^2}$ and $f_B = 37\,000\,\mathrm{meV^2}$. The same fitting procedure has been carried out with the µR spectrum obtained from the GaN/$\mathrm{Al_{0.08}Ga_{0.92}N}$/GaN heterostructure, keeping the same excitonic parameters for the excitons in the bottom thick GaN layer. The result is shown as a blue line in Figure \ref{Fig6}\textcolor{blue}{(b)}, with the excitonic transition energies shifted by a few meV to match the experimental results: $\mathrm{E(X_A^{n=1})}$ = 3494.3 meV and $\mathrm{E(X_B^{n=1})}$ = 3503.3 meV. However, this calculation does not reproduce the shoulder observed on the $A$ excitonic transition or the broadening of the $B$ excitonic transition given that the excitonic contribution of the top GaN surface layer was voluntary not introduced in this first simulation.\par
To improve the fit, we added the excitonic contribution of the thin GaN surface layer (red line in Figure \ref{Fig6}\textcolor{blue}{(b)}). The obtained excitonic parameters are as follows: $\mathrm{E(X_A^{n=1})}$ = 3494.8 meV; $\mathrm{E(X_B^{n=1})}$ = 3503.9 meV; $f_A = 1\, 800\,\mathrm{meV^2}$; $f_B = 1\,700\,\mathrm{meV^2}$; $\gamma_A = 0.5$ meV; $\gamma_B = 0.4$ meV; $\sigma_A = 0.3$ meV; and $\sigma_B = 0.7$ meV. We observe slight higher excitonic energies in the surface layer, consistent with the photoluminescence results shown in Figure \ref{Fig1}, and indicating a slight additionnal strain in the surface layer. Most importantly, the reduction of approximately 95\% of the oscillator strength of these excitons confirms that the thickness where the built-in electric field is sufficiently weak to maintain excitons is on the order of few nanometers, consistent with simulations presented in Figure \ref{Fig4}. 

\section{\label{sec:level6}Discussion}
The TRPL results presented in the previous section demonstrate that optical injection of free carriers into the heterostructure screens the electric field in the GaN layer, resulting in an increase of the excitonic lifetime. The modification of the heterostructure's band diagram and of the physical properties of excitons is reflected in changes of the photoluminescence spectra as a function of the excitation intensity, as shown in Figure \ref{Fig3}.\par
To describe this evolution, it is necessary to determine the carriers density optically injected using the Q-switched laser at 266 nm. This carrier density $n$ was determined graphically by comparing its evolution in pulsed and quasi-continuous regimes. When the GaN layer is excited with the Q-switched laser, the characteristic recombination times in GaN are shorter than the laser pulse duration ($\tau < t_p=400$ ps). In this case the quasi-continuous regime approximation is valid, and the carrier density can be expressed as:
\begin{equation}
n=G\tau (n)=\frac{(1-R)I\tau (n)}{h\nu d};
\end{equation}  
where $R$ is the reflectance at the vacuum/GaN interface ($\sim$ 0.7 for excitation at an incidence angle of 64° by taking into account the TE polarization of the laser). As in the pulsed regime, a uniform injection across the layer's thickness ($d=150$ nm) is assumed. For each optical intensity, the injected carrier density is deduced graphically from the intersection between the interpolation of the experimental points obtain in the pulsed regime (black line) and the $n(\tau)$ curve plotted for the corresponding intensity of the Q-switched laser (colored lines).\par
The photoluminescence intensity also depends on the exciton lifetime in the material. In the presence of an electric field, the exciton lifetime $\tau$ is determined by the radiative lifetime $\tau_r$, the non-radiative lifetime $\tau_{nr}$, and the dissociation time of the electron-hole pairs $\tau_d$, such that: $\tau^{-1}=\tau_r^{-1}+\tau_{nr}^{-1}+\tau_d^{-1}$. Thus, the photoluminescence intensity $i_{\mathrm{PL}}$ can be expressed as:\par
\begin{equation}\label{eq6}
i_{\mathrm{PL}}=\frac{n}{\tau_r}=\frac{G\tau}{\tau_r}.
\end{equation}
\begin{figure}
	\includegraphics[width=\columnwidth]{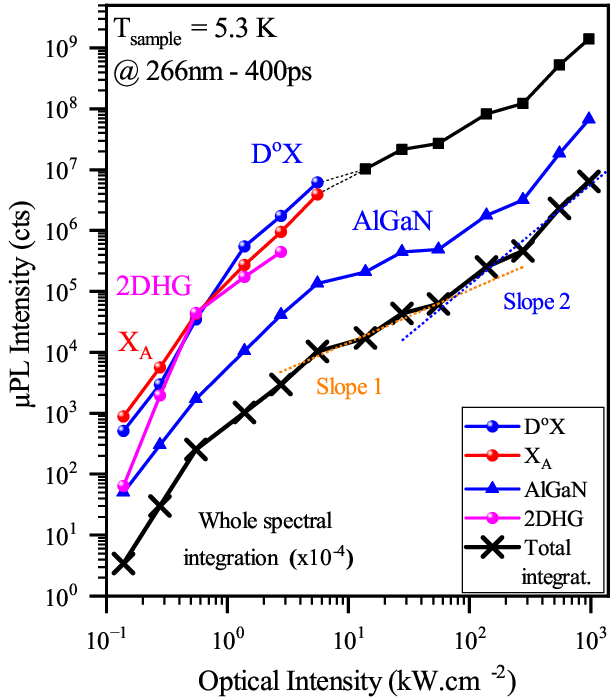}
	\caption{\label{Fig7} The evolution of the peak intensities for the different transitions observed in the luminescence spectra of the GaN/$\mathrm{Al_{0.08}Ga_{0.92}N}$/GaN heterostructure as a function of excitation intensity using the Q-switched laser at 266 nm is shown. The red line corresponds to the free A exciton, the blue circles represent the donor-bound exciton, and the blue triangles correspond to the $\mathrm{Al_{0.08}Ga_{0.92}N}$ transition. The pink line represents the 2DHG. The black squares correspond to the maximum intensity of GaN luminescence region. The integrated intensity over the entire luminescence spectrum (from 3.4 to 3.7 eV) is represented by black crosses, which have been vertically shifted downward for better visibility.}
\end{figure} 
Figure \ref{Fig7} shows the evolution of photoluminescence intensity of various emission lines as a function of laser excitation intensity. The GaN excitonic transitions and the $\mathrm{Al_{0.08}Ga_{0.92}N}$ luminescence follow a similar trend across the entire range of the investigated intensities. At low excitation intensities (from 150 $\mathrm{W\cdot cm^{-2}}$ to 12 $\mathrm{kW\cdot cm^{-2}}$), the intensity of the 2DHG peak increases sharply before decreasing and disappearing. Additionally, the inversion of the maxima between the free exciton peak $\mathrm{X_A}$ and the donor-bound exciton peak $\mathrm{D^\circ X}$ is observed, as explained earlier by the gradual screening of the internal electric field by the free carriers. The evolution of the integrated intensity over the entire luminescence spectrum (black crosses in Figure \ref{Fig7}) is explained by the screening of the electric field: the observed gradual decrease in the slope is associated to an increase of excitonic lifetime $\tau$, due to the increasing of dissociation time $\tau_d$ with the screening of built-in electric electric field, consistent with \textcolor{blue}{eq.}\ref{eq6}.\par
From 12 $\mathrm{kW\cdot cm^{-2}}$ to 55 $\mathrm{kW\cdot cm^{-2}}$, the intensity evolves linearly, indicating that the electric field is fully screened and the carrier lifetime is constant. The slope of 1 is characteristic of monomolecular recombination processes, such as excitonic transitions, where the intensity is proportional to the excitation density: $i_{\mathrm{PL}} \propto I$. Beyond 55 $\mathrm{kW\cdot cm^{-2}}$, the slope increases to 2, suggesting a shift in the recombination processes within the layers. This indicates a transition from monomolecular to bimolecular processes, where the intensity evolves quadratically with the excitation intensity: $i_{\mathrm{PL}} \propto I^2$.\par
The transition from excitonic recombination (monomolecular) to band-to-band recombination (bimolecular) indicates the screening of the Coulomb interaction between electrons and holes that form excitons, leading to the determination of the Mott density. The change in slope is observed for a carrier density of $n=4.6\times 10^{17}\, \mathrm{cm^{-3}}$. Since a portion of the optically injected carriers are used to screen the built-in electric field ($n_{\mathrm{scr}}=3.0\times 10^{16}\, \mathrm{cm^{-3}}$, corresponding to the first spectrum of Fig.\ref{Fig3} for which the 2DHG is no longer visible), their contribution must be subtracted from the Mott density, yielding a value of $n_{\mathrm{Mott}}=n-n_{\mathrm{scr}}=4.3\times 10^{17}\, \mathrm{cm^{-3}}$. This result aligns with recent literature findings \cite{Mallet_2022} for an electronic temperature of $T_e =$ 134 K, which report $n_{\mathrm{Mott}}=3.0\times 10^{17}\, \mathrm{cm^{-3}}$. Taking into account the determination of optically injected carriers densities and the experimental results presented in Fig.\ref{Fig3} and Fig.\ref{Fig7}, we estimate uncertainties of $2.0\times 10^{16}\, \mathrm{cm^{-3}}$ and $3.0\times 10^{17}\, \mathrm{cm^{-3}}$ for the density required to screen the built-in electric field and the Mott density, respectively.\par
Above the Mott density, peaks are still observable near the energies of free and bound excitons. Two hypothesis may be suggested: (i) the excitation is not homogeneous through the 150 nm thick GaN layer thus the Mott density is not reached everywhere or (ii) photons reemitted from the buffer GaN layer are amplified in the upper layer where gain is obtained in this energy range due to the bandgap renormalization. These peaks may be interpreted as amplified spontaneous emission (ASE). This second hypothesis agrees with the 0.5 meV red-shift of excitonic peaks at high excitation intensity observed on the luminescence spectra (Figure \ref{Fig3}) and with the results from the µR analysis (Figure \ref{Fig6}).

\section{\label{sec:level7}Conclusion}
In summary, the study of the GaN/$\mathrm{Al_{0.08}Ga_{0.92}N}$/GaN heterostructure has demonstrated the presence of excitons in the thin GaN layer (150 nm) using different experimental methods: (i) The analysis of luminescence spectra reveals the signal from the free and bound excitons in the top GaN layer when excited above the $\mathrm{Al_{0.08}Ga_{0.92}N}$ bandgap but also a weaker and slightly shifted signal originating from de $A$ exciton in the buffer GaN layer when excited below the $\mathrm{Al_{0.08}Ga_{0.92}N}$ bandgap; (ii) TRPL measurements using various excitation sources reveal a biexponential decay of the $A$ exciton with two characteristic times (18 ps and 70 ps) when the excitation wavelength is below the $\mathrm{Al_{0.08}Ga_{0.92}N}$ bandgap with a short decay time originating from the GaN thin layer and a longer one corresponding to excitons in the buffer layer; (iii) The study of photoluminescence spectra as a function of optically injected carrier density using a Q-switched laser at 266 nm demonstrates the presence of slightly redshifted excitonic transitions even beyond the Mott density in the surface GaN layer, with the high intensity of these peaks being explained by the amplified spontaneous emission (ASE) of photons reemitted from the buffer GaN layer in the surface GaN layer caused by the bandgap renormalization. (iv) µR measurements supported with numerical simulation highlight the contributions of two layers with excitons separated by an energy of 0.5 meV and with very low oscillator strength in the top GaN layer.\par 
This study also demonstrates the influence of the built-in electric field on free and bound excitons in the surface GaN layer and its impact on the heterostructure's luminescence. We calculated through the self-consistent resolution of the Schrödinger and Poisson equations that the built-in electric field is sufficient to ionize donors across a significant portion of the GaN surface layer. However, the inversion of the field direction defines a region where excitons can exist and donors are partially ionized. These findings are experimentally confirmed by a change in the ratio between the peak intensities of free and bound excitons: the donor bound exciton peak has a lower intensity at low excitation intensity but the ratio reverse as the excitation intensity increase. This behavior is observed in the same spectra as the 2DHG, indicating a dependance on the electric field. Moreover, the reduction of the exciton oscillator strength, determined through reflectivity fitting, further confirms that excitons exist within a region of only a few nanometers in the GaN surface layer. Dynamically, we also showed that the optical injection of free carriers screens the negative surface charge density at the GaN/$\mathrm{Al_{0.08}Ga_{0.92}N}$ interface through the accumulation of free holes in the 2DHG. The carrier density required to screen the built-in electric field was estimated to be $n_{\mathrm{scr}}=(3\pm2)\times 10^{16}\, \mathrm{cm^{-3}}$, at which point the transition associated with the 2DHG is no longer observed in the luminescence spectra. This screening is accompanied by an increase in the exciton lifetime, as measured by TRPL, due to the reduced dissociation rate of free excitons under the influence of the built-in electric field.\par 
Finally, we propose an estimation of the Mott density in the surface GaN layer of $n_{\mathrm{Mott}}=(4\pm3)\times 10^{17} \, \mathrm{cm^{-3}}$, which is similar to recent literature values.\par 
In conclusion, it is essential to account for the polar properties of nitride materials when studying also the optical properties of thick nitride layers (i.e. for thicknesses beyond those leading to quantum confinement). Despite high built-in electric fields, excitons are not completly dissociated and contribute to the luminescence of the heterostructure. Proper consideration of electric field screening is crucial to improve the thresholds of laser devices based on polar nitride materials.
\section{\label{sec:level8}Acknowledgments}
The authors acknowledge fundings from the French National Research Agency (ANR-21-CE24-0019-NEWAVE). We also thanks C2N, member of RENATECH, the french national network of large micro-nanofabrication facilities, for technological processes on our samples. We acknowledge support from GANEXT (ANR-11-LABX-0014); GANEXT belongs to the publicly funded ‘Investissements d’Avenir’ program managed by the Agence Nationale de la Recherche (ANR), France.

\end{document}